\documentclass[
 reprint,
 superscriptaddress,
 amsmath,amssymb,
 floatfix,
 prb,
 twocolumn
]{revtex4-1}

\usepackage[T1]{fontenc}
\usepackage[english]{babel}

\usepackage{amsmath}
\usepackage{graphicx}
\usepackage{epstopdf}
\usepackage{ulem}
\usepackage{upgreek}
\usepackage{bm}
\usepackage{lipsum}%
\usepackage{textcomp}
\usepackage{ifthen}
\usepackage[cmyk,dvipsnames]{xcolor}
\usepackage[colorlinks = true,
						citecolor = blue,
						linkcolor = red,
						urlcolor = blue]{hyperref}

\newcommand{\tf}[1]{{\textbf{#1}}}
\newcommand{\mr}[1]{\mathrm{#1}} 
\newcommand{\mf}[1]{\mathbf{#1}}  

\newcommand{\PdfccFeIr}[1]{Pd-fcc/Fe/Ir}

    \makeatletter
    \renewcommand*{\@fnsymbol}[1]{\ensuremath{\ifcase#1\or \dagger\or *\or \ddagger\or
        \mathsection\or \mathparagraph\or \|\or **\or \dagger\dagger
        \or \ddagger\ddagger \else\@ctrerr\fi}}
    \makeatother
\begin{document}

\title{Eigenmode following for direct entropy calculation and characterization of magnetic systems}

\author{Stephan von Malottki}
\affiliation{Institute of Theoretical Physics and Astrophysics, University of Kiel, Leibnizstrasse 15, 24098 Kiel, Germany}
\affiliation{Science Institute, University of Iceland, 107 Reykjav\'ik, Iceland}
\affiliation{Thayer School of Engineering, Dartmouth College, 15 Thayer Dr, Hanover, New Hampshire, USA}
\affiliation{UCLouvain, Institute of Condensed Matter and Nanosciences (IMCN), Chemin des Étoiles 8, Louvain-la-Neuve 1348, Belgium}

\author{Moritz A. Goerzen}
\affiliation{Institute of Theoretical Physics and Astrophysics, University of Kiel, Leibnizstrasse 15, 24098 Kiel, Germany}
\affiliation{CEMES, Universit\'e de Toulouse, CNRS, 29 rue Jeanne Marvig, F-31055 Toulouse, France}

\author{Hendrik Schrautzer}
\affiliation{Science Institute, University of Iceland, 107 Reykjav\'ik, Iceland}
\affiliation{Institute of Theoretical Physics and Astrophysics, University of Kiel, Leibnizstrasse 15, 24098 Kiel, Germany}

\author{Pavel F. Bessarab}
\affiliation{Department of Physics and Electrical Engineering, Linnaeus University, SE-39231 Kalmar, Sweden}
\affiliation{Science Institute, University of Iceland, 107 Reykjav\'ik, Iceland}

\author{Stefan Heinze}
\affiliation{Institute of Theoretical Physics and Astrophysics, University of Kiel, Leibnizstrasse 15, 24098 Kiel, Germany}
\affiliation{Kiel Nano, Surface, and Interface Science (KiNSIS), University of Kiel, Germany}



\maketitle

%-------------------------------------
% ABSTRACT
%-------------------------------------
\textbf{We present an eigenmode following method (EMF) that has been developed for the numerical scanning of the potential energy surface and the direct calculation of entropy and pre-exponential factors of Arrhenius-like transition rates in the framework of transition state theory. In contrast to other methods, we do not use EMF to move "uphill" or "downhill" in energy to find stationary points, but to obtain energy curves. By numerical integration of the Boltzmann factor along these curves, the partition functions of the followed eigenmodes can be calculated without making assumptions of the form of the energy curves. The EMF method is computationally more elaborate than traditional approaches as it requires iterative updates of the respective eigenpair spectrum. In order to mitigate this increase in computational cost, the EMF method can be combined with other approaches like the harmonic approximation for all eigenmodes but the softest which typically require the most accurate modelling. In this work we first introduce the general theoretical background and algorithm of the EMF method before providing test calculations and relevant use-cases in the framework of atomistic spin simulations with a focus on magnetic skyrmions collapse.}

%-------------------------------------
\section*{Introduction}
%-------------------------------------
%
Various techniques to fully or partially explore potential energy surfaces (PES) have been developed in the past in order to study the stability and reactivity of solids, molecules and systems with multiple degrees of freedom in general \cite{schlegel2003exploring, doye1997surveying, steinmetzer2021pysisyphus, Nichols1990walking}. Some "walk" on the PES to find local minima and transition states \cite{baker1986algorithm, banerjee1985search, steinmetzer2021pysisyphus, olsen2004comparison, cerjan1981on, doye1997surveying, Nichols1990walking, henkelman1999dimer, muller2018duplication}, while others, for example the nudged elastic band method (NEB), energetically relax a reaction path to the minimum energy path (MEP) between an initial and final state, providing information about reaction mechanisms connecting these states \cite{jonsson1998nudged, sheppard2012generalized, henkelman2000improved}. In combination with statistical methods such as transition state theory (TST) \cite{wigner1938transition, evans1935some, eyring1931ueber, glasstone1941theory} and related methods \cite{langer1969statistical}, knowledge of these points on the PES allows for the calculation of energy barriers and transition rates, typically written in the form of an Arrhenius law:
\begin{equation}
    \nu = \nu_0~ e^{-\beta \Delta E}.
    \label{Eq:arrhenius_law}
\end{equation}
Here, the transition rate, $\nu$, is given by the pre-exponential factor, $\nu_0$, the inverse of the thermal energy, $\beta = 1 / k_\mathrm{B} T$, and the energy barrier between the two states, $\Delta E$. The energy barrier is defined as the energy difference between transition state and initial state and can be obtained for example via the climbing image-NEB method (CI-NEB) \cite{henkelman2000climbing}. The pre-exponential factor, $\nu_0$, on the other hand, contains the dynamic and entropic components of the transition rate and is often more difficult to estimate \cite{beste2010one, von2019skyrmion, desplat2018thermal}. 

In magnetism, variants of NEB have been used to study spin flip events \cite{dittrich2002path}, domain wall switching \cite{dittrich2002path} and vortex switching \cite{thiaville2003micromagnetic, krone2010magnetization}, often in the context of thermal stability of magnetic bits \cite{dittrich2003energy}. However, it was the discovery of magnetic skyrmions in 2009 \cite{muehlbauer2009skyrmion}, small topological, whirl-like structures, that motivated more sophisticated theoretical investigations \cite{back20202020, tokura2020magnetic, gobel2021beyond, heinze2011spontaneous}. By applying the geodesic nudged elastic band method (GNEB) \cite{bessarab2015method}, a variant of NEB that takes the constraint length of magnetic moments into account, the radial \cite{bessarab2015method, lobanov2016mechanism, desplat2018thermal, von2017enhanced}, chimera \cite{desplat2019paths, meyer2019isolated} and escape mechanisms \cite{bessarab2018lifetime} of skyrmion decay have been found and enabled the calculation of their corresponding energy barrier, $\Delta E$. More recently, the first two mechanisms have also been identified experimentally \cite{muckel2021experimental}. Since the introduction of GNEB, the mechanisms and energy barriers of antiskyrmions \cite{bessarab2015method, von2017enhanced, goerzen2023lifetime, hoffmann2021skyrmion}, skyrmions in antiferromagnets \cite{bessarab2019stability}, with higher-order exchange interactions \cite{paul2020role}, in multilayer systems \cite{schrautzer2022, hoffmann2021skyrmion, hoffmann2020atomistic}, skyrmioniums \cite{hagemeister2018controlled}, 
higher-order skyrmions \cite{desplat2019paths}
and skyrmions in 3D systems \cite{leishman2020topological, birch2021topological} have been investigated. 

Further, the introduction of TST in harmonic approximation (HTST) to atomistic spin simulations \cite{bessarab2012harmonic, desplat2018thermal, von2019skyrmion, bessarab2018lifetime} enabled the calculation of the pre-exponential factor and thus, mean lifetimes of skyrmionic textures in model systems \cite{varentcova2020toward, desplat2018thermal, desplat2019paths, goerzen2022atomistic} as well as realistic, DFT-parameterized ultrathin film systems \cite{von2019skyrmion, goerzen2023lifetime, hoffmann2020atomistic, haldar2018first}. The relevance of taking the prefactor into account was demonstrated experimentally in the chiral magnet Fe${}_{1-x}$Co${}_x$Si by a dramatic change in prefactor by over 30 orders of magnitude over a small variation of magnetic field \cite{wild2017entropy}. It was possible to calculate this entropic stabilisation effect for other materials via HTST as well and link it to the dependency of the prefactor on a few localised skyrmion eigenmodes \cite{desplat2018thermal, von2019skyrmion, malottki2021stability, hoffmann2021skyrmion}.

We use the term eigenmode for a pattern of movement initially defined by the eigenvectors of the Hessian matrix. However, not all skyrmion eigenmodes can be modelled adequately in harmonic approximation. Translation and rotational modes of skyrmions and antiskyrmions are often better treated in zero-mode, also called Goldstone mode \cite{bessarab2018lifetime, von2019skyrmion}, approximation, which assumes the energy to be constant along the eigenmode, while the rest of the spectrum is still treated in harmonic approximation \cite{bessarab2018lifetime, von2019skyrmion, desplat2019paths}. Some magnetic states exhibit eigenmodes that do not fall into any of these two categories, limiting the applicability and accuracy of HTST to skyrmionics, especially for varying temperature regimes. For instance, the rotational modes of the Chimera saddle point structure as well as the antiskyrmion groundstate prohibited the calculation of mean lifetimes in some studies \cite{desplat2019paths, meyer2019isolated}. The community was partially circumventing these issues by performing direct Langevin dynamics \cite{rohart2016path} or Monte Carlo \cite{hagemeister2015stability} simulations, but these methods are limited by the rare event problem of skyrmions with high mean lifetime which in turn can partially be mitigated by sophisticated forward-flux calculations \cite{desplat2020path}.

The limitations of the harmonic approximation are well-known in other fields \cite{collinge2020effect, jorgensen2017adsorbate, amsler2021anharmonic, sauer2024future}. They can stem from the energy curves along eigenmodes simply not being shaped parabolic around the expansion point \cite{collinge2020effect, beste2010one}, the coupling between several eigenmodes \cite{collinge2020effect, beste2010one} or a combination of both. In this work, we do focus on errors of the first kind and neglect the second. By applying HTST in classical approximation, we also neglect quantum mechanical effects such as tunneling events. However, for skyrmion collapse in ultrathin films such as Pd/Fe/Ir(111) cite{romming2013writing,muckel2021experimental},
these effects are thought to be small \cite{vlasov2020magnetic}. 
A common way to improve the description of the potential energy curve is to perform a fit with a better suited formula. Examples are the free translator/rotor models \cite{li2018zeolite}, which are other names for the zero-mode approximation, a periodic potential, the so-called "hindered translator/rotator" models \cite{sprowl2016hindered} or polynomials of third or fourth order \cite{beste2010one, piccini2013quantum, piccini2014effect}. However, it has been shown that these approximations can still fail in describing the energy along the curve adequately, leading to large errors in the calculated partition functions and consequently the entropies \cite{jorgensen2017adsorbate, beste2010one, amsler2021anharmonic}. The promising "complete potential energy sampling" (CPES) method performs a direct numerical calculation of the entropy without relying on a pre-defined form of the energy curve \cite{jorgensen2017adsorbate}. The CPES method, however, requires explicit knowledge of several relevant points on the PES and a reasonable interpolation between them, which currently limits its generality \cite{jorgensen2017adsorbate, collinge2020effect}. 

\begin{figure}
    \centering
    \includegraphics[width=78mm]{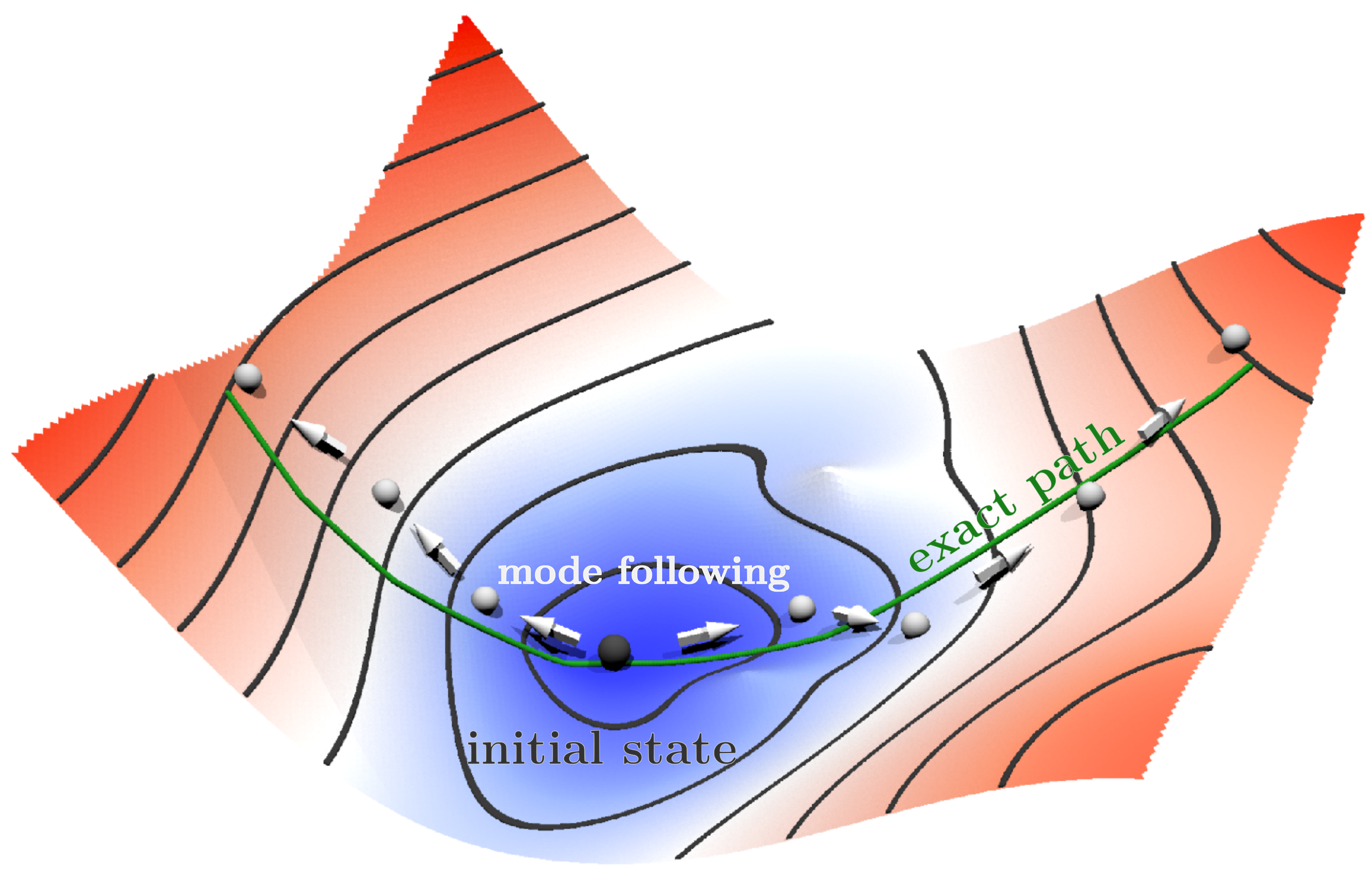}
    \caption{%
    \textbf{Illustration of the eigenmode following method.} 
    The shape and colour of the 2D function illustrate an arbitrary potential energy surface with a local energy minima. From an initial state (black sphere), the eigenmode following method translates the system iteratively in small steps along the followed eigenvector, yielding an approximated path and energy curve of the eigenmode. The current eigenvectors (arrows) of the current state (white spheres) are re-calculated each step. %
    }%
    \label{fig:illustration_mfm}
\end{figure}

Here, we present an eigenmode following method (EMF) for direct entropy calculation which combines the numerical evaluation of entropy by the sum over states formula \cite{beste2010one} (Eq. \ref{Eq:cases_phi}, numerical) with eigenmode following. In contrast to various well-established techniques \cite{muller2018duplication, plasencia2017improved, tsai1993use, olsen2004comparison}% 
, we do not use EMF to find stationary points on the PES but to classify individual eigenmodes and obtain their energy curves with numerical accuracy without approximating their form with any kind of formula. While methods proposed by Beste \cite{beste2010one} as well as by Sauer and coworkers \cite{piccini2013quantum, piccini2014effect} perform small step EMF in linear response to fit higher-order polynomials to the obtained local energy curve, our EMF goes beyond this approach by iteratively updating the followed eigenvector as illustrated in Fig. \ref{fig:illustration_mfm}. This allows accurate EMF over much larger distances, and thus, the direct numerical evaluation of the entropy. 

In the following, the theoretical background and algorithm of EMF for direct entropy calculation is introduced, followed by test calculations and use cases for magnetic skyrmions. Adequate application of TST for the calculation of mean skyrmion lifetimes has not been possible for these use-cases prior to the development of the EMF method. All three examples have been published elsewhere and here we just summarize how the EMF method has been applied in order to treat their anharmonic eigenmodes. While we limit ourselves to the framework of quasi-classical atomistic spin simulations, we do believe that the idea of our EMF can be of use other fields and incorporated into more complex and quantum mechanical approaches such as density functional theory.

%-------------------------------------
\section*{Eigenmode following for transition state theory}
%-------------------------------------
For the sake of clarity, we just give a brief introduction to harmonic transition state theory here and refer to other publications with a more general and extensive explanation of the method \cite{wigner1938transition, evans1935some, eyring1931ueber, glasstone1941theory, bessarab2012harmonic,varentcova2020toward, malottki2021stability}.
At the very basic, the main task in TST is to calculate the flux through the dividing surface S, which can be expressed by the following integral:
\begin{equation}
    \nu = \int_S{\mathrm{e}^{-\beta E(\mathbf{Q})}~ v_{\perp}^+ (\mathbf{Q})~ \mathrm{d}\mathbf{Q}}.
\end{equation}
Each state of the system is described by the generalised coordinates, $Q = q_1, ..., q_N$ and the component of the velocity perpendicular to S in positive direction of the reaction is given by $v_{\perp}^+$.

By applying the independent mode approximation \cite{beste2010one}, the pre-exponential factor of the Arrhenius law (see Eq. \ref{Eq:arrhenius_law}) can be written in a general form as
\begin{equation}
    \nu_0 = \frac{\lambda}{\sqrt{2\pi\beta}}\frac{\prod_{n=2}^{3N} \Phi_n^{\text{SP}}}{\prod_{n=1}^{3N}\Phi_n^{\text{I}}}.
\label{Eq:prefactor_general}
\end{equation}
Here, $\lambda$ contains the dynamic contribution of the transition (see Ref. \cite{varentcova2020toward} for more details) and the ratio of products consists of the partition functions of the eigenmodes of the initial and saddle point (SP) states, $\Phi_n^{\text{I}}$ and $\Phi_n^{\text{SP}}$, respectively. % 
The number of atoms is given by $N$. The indices of the eigenmodes are ordered with increasing eigenvalues of the respective Hessian matrix. %
Note, that the partition function of the first eigenmode of the SP, also referred to as the unstable mode, is excluded from the product, since states along this mode do not lie within the dividing surface.

Due to the decoupling of the eigenmodes, i. e. the independent mode approximation, their partition functions can be calculated individually via the line integral
\begin{equation}
        \Phi_n = \int_{-\infty}^{\infty} e^{-\frac{\beta}{2}[E(q_n)-E_0]}~ \mathrm{d}q_n,
        \label{Eq:phi_integral}
\end{equation}
where $q_n$ denotes the value of the eigenvector $\mathbf{q}_n$, whose orientation can change during the mode following process. $E(q_n)$ is the energy curve along $q_n$ and $E_0$ is the energy of the state at $q_n = 0$. %
The most common approximations of the energies $E(q_n)$ follow a Taylor expansion \cite{beste2010one}:
\begin{equation}
    E ( q_n)~ \approx \underbrace{x_0}_{\substack{\text{zero} \\ \text{mode}}} +~ \underbrace{x_1 q_n}_{ = 0}~ + \underbrace{x_2 q_n^2}_{\text{harmonic}} +~ \underbrace{x_3 q_n^3~ +~ x_4 q_n^4.}_{\substack{\text{higher order} \\ \text{corrections}}}
\end{equation}
Consequently, the approximations are named after the highest expansion parameter $x \neq 0$. Stationary points are defined by their first derivative equalling zero and, thus, the first order element vanishes. 

In this work, we compare the partition functions obtained via three methods: Zero mode approximation, harmonic approximation and our EMF for direct calculation of the entropy, which is directly related to the partition functions $\Phi_n$ by
\begin{equation}
    S = k_{\text{B}}\left(\frac{N_h}{2} + \sum_{n=1}^{3N}\ln\Phi_n\right)
    \label{eq:entropy}
\end{equation}
where $N_h$ is the number of harmonic degrees of freedom. The corresponding formula for the partition function of the $n$-th eigenmode of the diagonalized Hessian matrix read:
\begin{equation}
    \arraycolsep=1.4pt\def\arraystretch{2.2}
    \begin{split}
    \Phi_n = \left\{\begin{array}{cc}
    \displaystyle V_n, & \text{zero mode}, \\
    \displaystyle\sqrt{2\pi(\beta\epsilon_n)^{-1}}, & \text{harmonic}, \\
    \displaystyle\sum_{k=1}^{K} \mr{e}^{-\beta \left[E\left( q_n^k\right)-E_0\right]} \Delta q_n^k, & \text{numerical.}
    \end{array}\right.
    \end{split}
    \label{Eq:cases_phi}
\end{equation}
The zero mode volume, $V_n$, is independent of energies and the occupation of states. It equals the uniform integral over $q_n$ and requires prior knowledge about it's symmetry. The partition function of the harmonic approximation depends on the thermal energy, $\beta$, and the eigenvalue of the $n$-th eigenmode, $\epsilon_n$, which represents the curvature of the PES at the expansion point, i.e. the investigated state. 

In the EMF calculation of entropy, the integral of Eq. \ref{Eq:phi_integral} is replaced by a finite sum of small steps, $k$, along the eigencoordinate $q_n$ with a step width of $\Delta q^k_{n}$. As illustrated in Fig. \ref{fig:illustration_mfm}, these steps are performed iteratively along the $n$-th eigenmode. For each step, the energy is evaluated and used to calculate the Boltzman factor which yields the local contribution to the partition function when multiplied by the step width. In contrast to other methods such as Ref. \cite{beste2010one, piccini2013quantum, piccini2014effect}, the EMF for direct entropy calculation does not require any explicit approximation of the PES and describes the energy curve with numerical precision as long as the mode tracking (see next section) is successful. Note that the EMF as well as the above mentioned approaches rely on the independent mode approximation and, thus, a reasonable amount of orthogonality of the eigenvectors along the integrated path. 

While each $\Phi_n$ can be calculated using a different approximation, it is the softest eigenmodes of a system that can have a strong effect on the overall transition rates. While showing only a small absolute error when modelled inadequately, their relative error can be dramatic due to the multiplication with the rest of the distribution function. It is a practical strategy to calculate the vast majority of $\Phi_n$ in harmonic approximation and use the more costly and accurate EMF for a few eigenmodes with the lowest eigenvalues.

\section*{Algorithm}
\begin{figure}
\includegraphics[width=58mm]{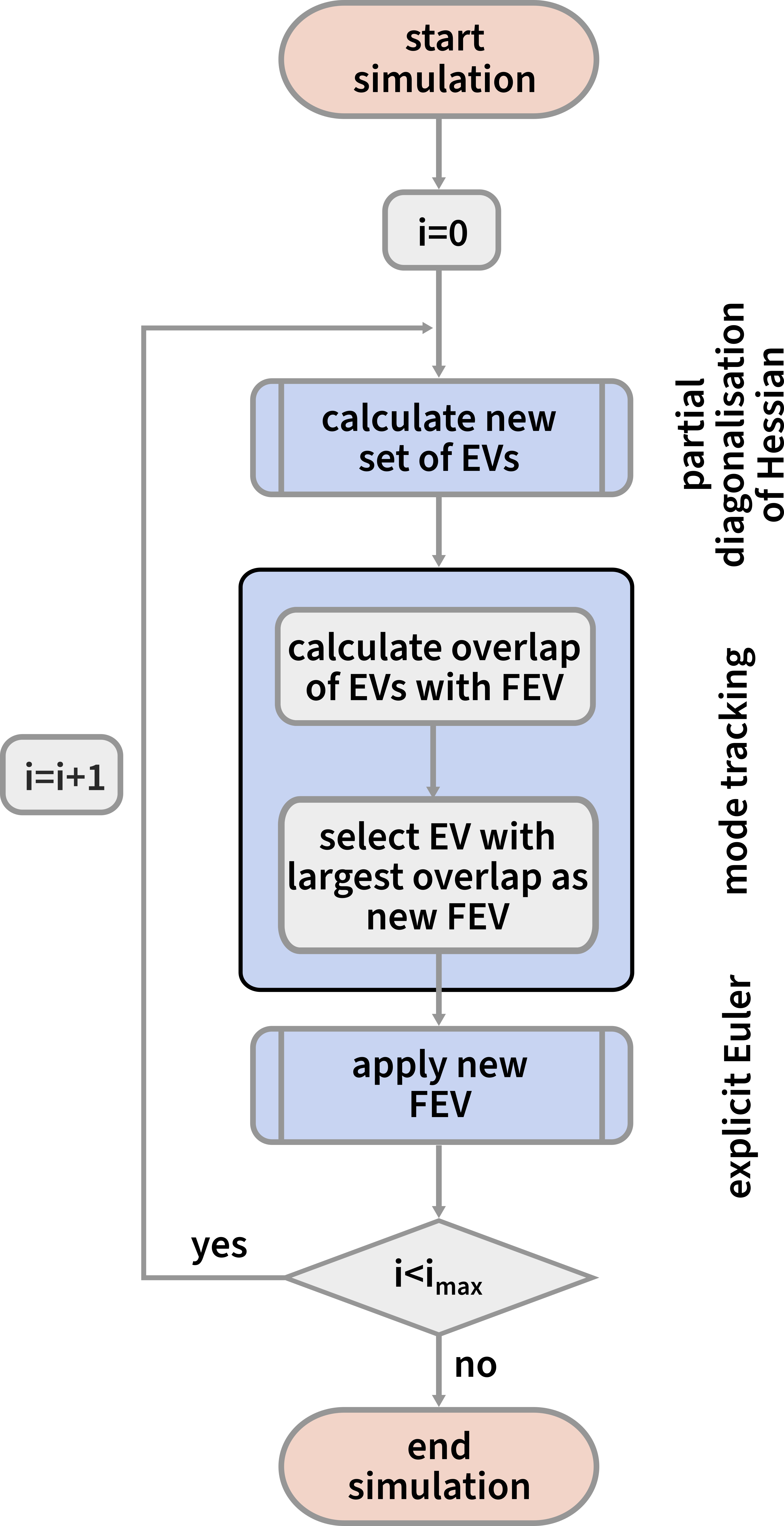}
\caption{\label{fig:algorithm_mode_following}
\textbf{Abstract flow chart of the eigenmode following method.} %
In the iterative loop, the three main subroutines are executed one after another: The partial or full diagonalisation of the Hessian which yields a new set of eigenvalues and eigenvectors (EVs) for the current state. Then the mode tracking, which calculates the overlap of the previously followed eigenvector (FEV), $v_i^F$, with the new set of EVs and than select the one with the largest overlap as the next followed eigenvector, $v_{i+1}^F$. Finally, the newly found $v_{i+1}^F$ is applied to the current state, $Q_i$, creating the next state, $Q_{i+1}$, as a basis for the next iteration cycle. The iteration stops when the maximum number of steps, a maximum distance of displacement or a certain energy is reached.%
}
\end{figure}
Starting from a selected system configuration, $Q_0$, on the PES, typically the initial or SP state, the EMF method iteratively displaces the system configurations, $Q_i$, in the direction of the selected eigenvectors, $v_i^F$, of the corresponding Hessian matrices. The followed eigenvector is updated at each step to allow for long range displacement, minimizing distortions along the path. The algorithm, sketched in Fig. \ref{fig:algorithm_mode_following}, consists of four primary steps:
\begin{enumerate}
    \item Calculation of the eigenvalues, $\{\lambda^m_i\}$, and eigenvectors, $\{v^m_i\}$, of the system configuration $Q_i$.
    \item Identify which eigenvector of $\{v^m_i\}$ is the most similar to $v_{i-1}^{F}$ (Mode tracking).
    \item Displacement of the current configuration, $Q_{i}$, along the selected eigenvector $v_i^F$.
    \item Evaluation of the energy of the new state $Q_{i+1}$.
\end{enumerate}

\begin{figure}
    \centering
    \includegraphics[width=78mm]{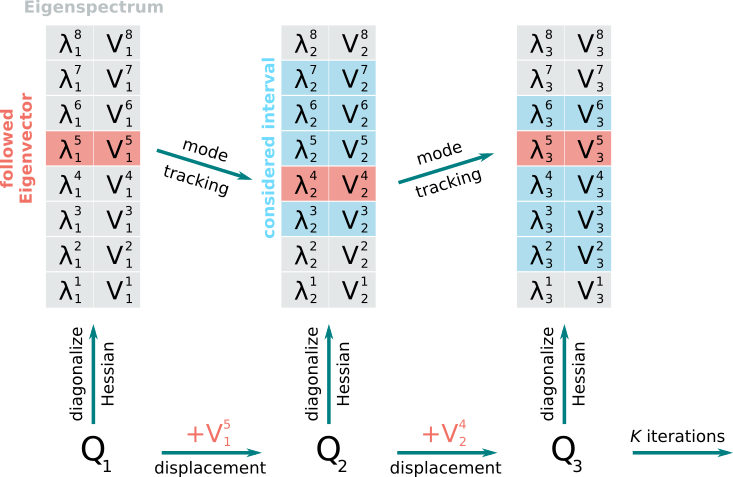}
    \caption{\textbf{Illustration of the Mode tracking and the considered interval of the eigenspectrum.} %
    An initial state $Q_1$ is diagonalised, yielding the ordered eigenspectrum with eigenvalues, $\{\lambda_1^m\}$, and their corresponding eigenvectors, $\{v_1^m\}$. In this example we assume that the fifths eigenmode should be followed, so $Q_1$ is displaced in the direction of $v_1^5$, creating the next state in the iteration, $Q_2$. This time, a certain interval centred around the previous selected eigenvector, here $v_1^5$, is chosen, here with two eigenpairs above and below index 5. In the second iteration, the eigenvector $V_2^4$ has maximum overlap with $V_1^5$ and is applied to $Q_2$ in order to calculate $Q_3$. This procedure is repeated for $K$ iterations.%
    }
    \label{fig:mode_tracking}
\end{figure}
Here, $m$ denotes the index of the eigenvalue in the ordered spectrum. In principle, the eigenpair spectra, $\{\lambda^m_i\}$ and $\{v^m_i\}$, can be obtained by direct diagonalisation of the Hessian matrix of the state $Q_i$. It consists of the second derivatives of the Hamiltonian with regard to the generalised coordinates, $q_i$. While this approach is straight forward, it is also often the computational bottle-neck in large systems. In most cases, it is only necessary to treat eigenmodes at the lower end of the spectrum with EMF as they are the softest and most likely eigenmodes to exhibit non parabolic energy curves. Therefore, just the lowest few eigenpairs need to be calculated and extremal eigensolvers can be utilized to significantly reduce the computational costs of the diagonalisation. 

In order to further reduce the computational cost, other, more case dependent optimizations can be applied. For example, an Hamiltonian without long range interactions can result in a sparse Hessian matrix, which is the case for our extended Heisenberg model (see Eq. \ref{eq:heisenberg_model}) that we apply in the next sections. This allows us to use a state-of-the-art extremal sparse eigensolver\cite{stewart2002krylov} implemented in the Intel MKL library. The Hessian matrix can also be approximated as for example by finite difference approximations using the first order information of the gradients \cite{henkelman1999dimer,sallermann2023}. This flexibility of the method allows the user to find a physically and computationally sound implementation that suits the studied system the most to make it as fast and accurate as possible.

When the new eigenvectors, $\{v^m_i\}$, have been obtained, the next task is to identify which of them corresponds to the selected eigenmode to follow. In our EMF method, this means the "mode tracking" algorithm has to determine which new eigenvector is the most similar to the followed eigenvector of the previous iteration, $v^F_{i-1}$. In the first iteration, $i=1$, the followed eigenvector, $v^F_1$, is selected by the user. In the subsequent iterations, the overlap of the new eigenvectors with the previous $v^F_{i-1}$ is calculated. In the framework of atomistic spin simulations the eigenvectors are $2N$ dimensional vector fields in tangent space and the overlap translates to simple scalar products of $v^m_i$ and $v^F_i$. 

In large systems it is impractical to calculate the overlap of $v^F_i$ with each individual eigenvector of the spectrum. Instead, we use the indexation of the eigenvectors in increasing eigenvalue order and the condition that the PES is continuously derivable twice. This means, that a sufficiently small step on the PES results in a small change in the eigenspectrum, and thus, only a limited shift of the eigenvector indexation. This allows us to consider only a small interval of eigenvector indices around the index of $v^F_{i-1}$ and to still be ensured to find the new eigenvector with maximum overlap. This procedure is illustrated in Fig. \ref{fig:mode_tracking}.

Note that mode tracking only works reliably when there are no hybridisations of eigenmodes in the considered interval of eigenvectors. If that is the case, the characters of the corresponding eigenmodes are changing and the concept of following an eigenmode breaks down. An example would be the compression of a magnetic skyrmion until it is too small to be compressed further and the corresponding eigenvector gets pushed far up the spectrum or even disappears entirely. Fortunately, this phenomenon is negligible for the numerical integration of the entropy (Eq. \ref{Eq:cases_phi}) when the energy is increasing quickly enough while following the eigenmode for the Boltzmann factor to decay before the breaking point of mode following is reached. 

After the new followed eigenvector, $v^F_i$, has been identified via mode tracking, it is applied to displace the current system configuration:
\begin{equation}
Q_i = Q_i + \Delta s~ v^F_i.
\label{eq:displacement}
\end{equation}
Here, $\Delta s$ is the step width. The displacement can be performed with different numerical recipes as for example with an explicit Runge-Kutta algorithm or the Velocity Projection Optimization (VPO) method \cite{bessarab2015method}. The choice might depend on the studied system and the computational framework. Here, we apply a simple direct Euler integration scheme which yields sufficient numerical stability and accuracy for the systems studied in this work. Note that for atomistic spin simulations the configuration space is a $2N$ dimensional hypersphere and displacements have to be performed as rotations in the tangent space as described for example in Ref. \cite{bessarab2015method, malottki2021stability}. 

%--------------------------------
\section*{EMF in atomistic spin simulations}
We have developed and tested the EMF method in the context of atomistic spin simulations, which are applying a quasi-classical atomistic approximation to describe the magnetic moment of solid states. For all test and use cases presented in this work, the interactions of the magnetic moments are given by the extended Heisenberg model:
\begin{equation}
    \begin{split}\label{eq:heisenberg_model}
        E &= -\sum_{i,j} J_{ij} (\mathbf{m}_i\cdot\mathbf{m}_j)-
        \sum_{i,j} D_{ij}(\hat{\mathbf{z}}\times\mathbf{r}_{ij})\cdot(\mathbf{m}_i\times\mathbf{m}_j) \\
        &-\sum_{i}K(\mathbf{m}_i\cdot\hat{\mathbf{z}})^2-\sum_{i}
        \mu_i (\mathbf{m}_i\cdot\mathbf{B})~.
    \end{split}
\end{equation}
It takes into account two-site exchange interactions beyond nearest neighbours, Dzyaloshinskii-Moriya interaction (DMI), magnetocrystalline anisotropy energy (MAE) and Zeeman interaction. Due to the constrained length of the magnetic moments, the configuration space changes from the general $3N$ to a $2N$ dimensional hypersphere with $N$ being the number of atoms or magnetic moments, respectively.

The limitation to this specific model and field of application is not intrinsic to the EMF method but reflects our own scientific focus and expertise. However, it does allow for a quasi-classical treatment of the state space and an explicit construction of a Hessian matrix which, due to the lack of dipole-dipole interaction, is sufficiently sparse for efficient sparse extremal eigensolver algorithms. In the following, two test cases of EMF are presented in order to illustrate and test the implementation. It is followed by three very recent examples of EMF use-cases in which the method opened the door to obtain new insights in skyrmionics.

%--------------------------------
\subsection*{Single spin: Comparison of partial partition functions}
\begin{figure}
\includegraphics[width=86mm]{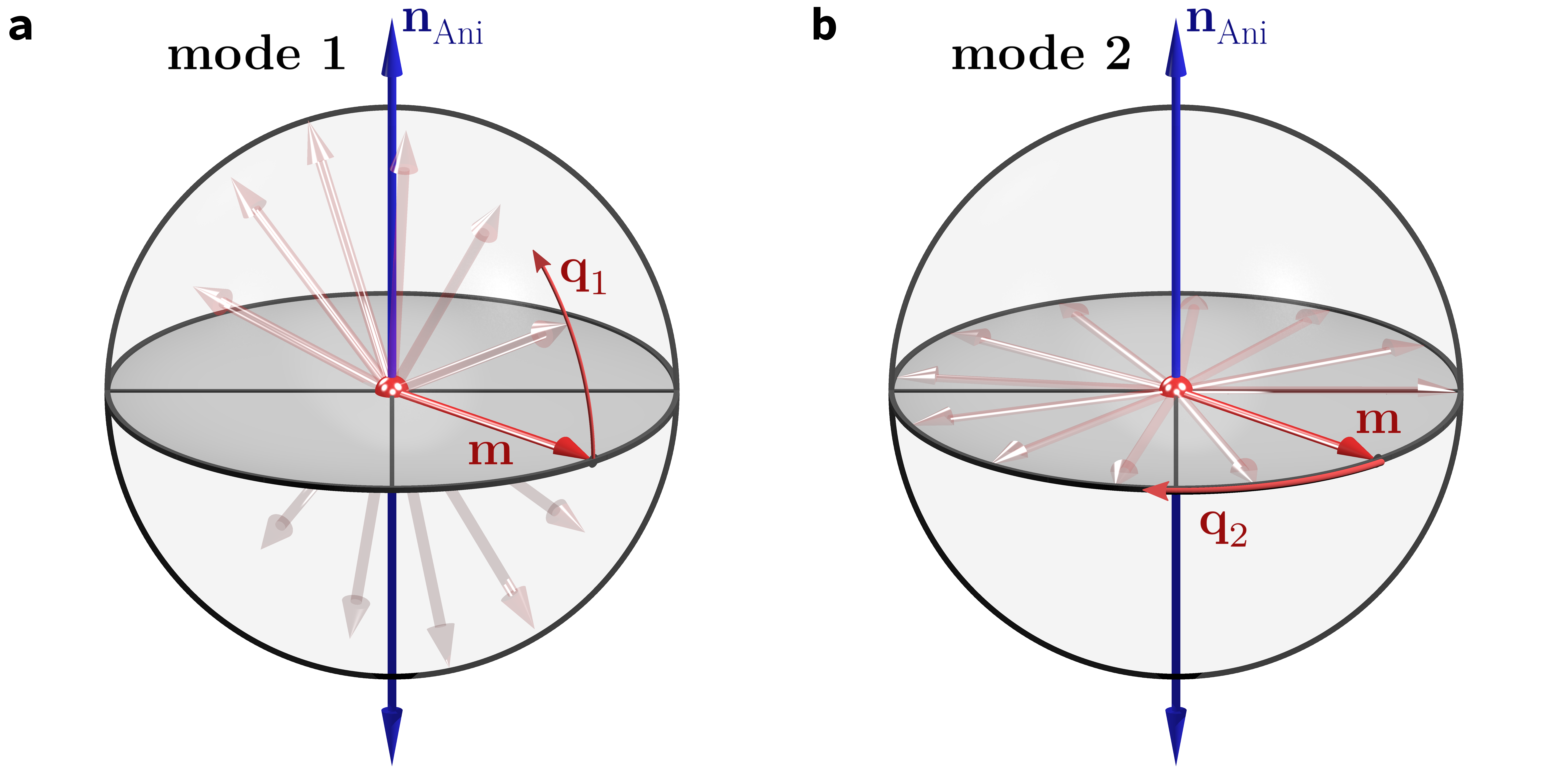}
\caption{\label{fig:single_spins}%
\tf{Illustration of the eigenmodes of a single spin system with uniaxial anisotropy in in-plane configuration.} %
The magnetic moments and the axis of the MAE are illustrated by red ($m$) and blue ($n_\text{Ani}$) arrows, respectively. The shaded red arrows indicate the movement of $m$ along the eigenmodes $i$ with the eigenvectors $q_i$. \textbf{a)} The magnetic moment rotates from pole to pole, gaining and losing energy in the process. The two-fold degenerate eigenmode of the spin in out-of-plane configuration looks similar, just with the initial spin orientation along $\hat{\mf{z}}$. \textbf{b)} The magnetic moment rotates in the equatorial plane, preserving its energy.} 
\end{figure}
Here, we consider a single magnetic moment in two configurations: Parallel and perpendicular to the uniaxial anisotropy, i. e. the $\hat{\mf{z}}$-axis. %
The diagonalisation of the $2\times2$ dimensional Hessian matrix in tangent space coordinates results in two eigenmodes. %
For a magnetic moment in easy axis direction, the two eigenmodes of the system are degenerate and correspond to a rotation from one pole over the equator towards the other pole. When assuming the magnetic moment to be oriented in the hard plane, which coincides with the SP of a pole-to-pole transition, the diagonalization yields the two eigenmodes illustrated in Fig.~\ref{fig:single_spins}. The first one is of the same nature as the two degenerate eigenmodes of the energy minimum. The second eigenmode describes a movement around the equator with constant energy.  

The partial partition functions $\Phi_p$, introduced in Eq.~(\ref{Eq:phi_integral}), equal the statistically occupation of states along the eigenmodes with index $p=1,2$. In this example, we assume a temperature of $T=10$~K and a MAE constant of $K=1$~meV. The corresponding eigenvalues of the degenerate eigenmode at the minimum are $\lambda_1^{\text{min}}=2$~meV and the two values of the SP modes are $\lambda_1^{\text{SP}}=-2$~meV and $\lambda_2^{\text{SP}}=0$~meV. Table \ref{Tab:phi_single_spin} shows the values of the partition functions in Eq.~(\ref{Eq:phi_integral}) determined via different techniques. The values of $\Phi_p$ are calculated for a full rotation over $\theta$ and $\phi$, respectively, and details on the calculations are given in the method section. The method coined "numeric" serves as a reference and corresponds to the direct numerical integration of the $\Phi_p$ integrals, made possible by the simple geometry of this test case.  

\begin{table}
\centering
\caption{Partition functions of both eigenmodes of the single magnetic moment parallel to the easy axis of the magnetocrystalline anisotropy, obtained with different methods: The harmonic approximation, zero-mode approximation, EMF method and the numerical exact result.}%
\begin{ruledtabular}
\begin{tabular}{lcccc}
$\Phi_p$ & harmonic & zero-mode & EMF & numeric\\ \hline
mode 1 min& $3.291$ & $6.283$ & $3.819$ & $3.819$\\
mode 1 SP& - & $1.969$ & $3.819$ & $3.819$\\
mode 2 SP& - & $1.969$ & $1.969$ & $1.969$ 
\end{tabular}
\end{ruledtabular}
\label{Tab:phi_single_spin}
\end{table}

For the eigenmode of the polar configuration, only the EMF method reproduces the value of the numeric method, while the harmonic approximation undercuts the partition function due to the anharmonicity of the energy curve. The zero-mode approximation, on the other hand, overestimates the occupation as it does neglect the increasing energy cost away from the easy axis. 

The first SP mode cannot be treated with harmonic approximation as it is an unstable mode and exhibits a negative eigenvalue. In the zero-mode approximation the energy is assumed to be constant and equal to the SP energy. This underestimates the occupation closer to the poles leading to a too small value. Again, only the EMF method reproduces the values of the numeric integration. For the second eigenmode of the SP, the zero-mode approximation is accurate as the energy does not change with varying orientations in the equator plane. The harmonic approximation cannot be applied due to the vanishing eigenvalue of that mode, which would lead to a diverging value of $\Phi_p$. The three other modes all agree well and yield the correct results.

Besides an intuitive illustration of the EMF method and its application, the above test calculations demonstrate that even in a simple one spin system, both, the harmonic and zero-mode approximation can fail to yield the partition functions of eigenmodes and thus their entropic contribution to stability properties while the EMF method produces correct results. 

\subsection*{Ferromagnet: Numerical limits of integration}
\begin{figure}[h!]
\includegraphics[width=84mm]{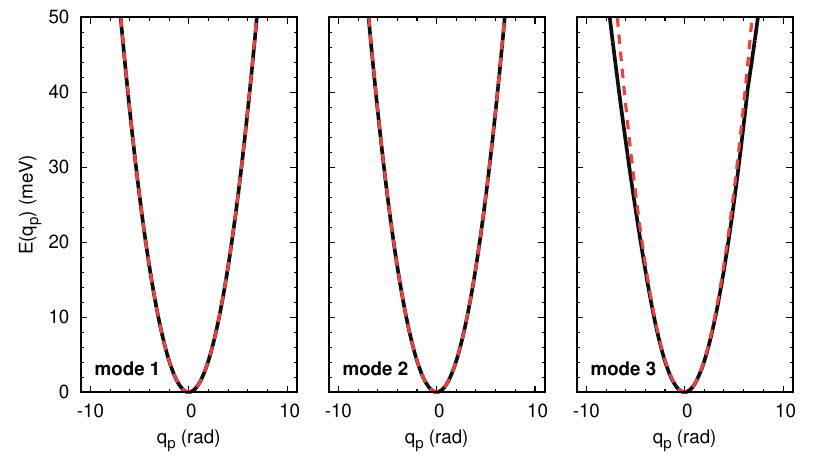}
\caption{\label{fig:fm_energy_dispersion}%
\textbf{Energy dispersion of magnon modes.}%
The lowest three eigenmodes of the ferromagnetic state were followed by EMF. As test system, the \PdfccFeIr{} parameters and a out-of-plane magnetic field of $4.0~T$ are chosen. The black and red lines correspond to EMF and harmonic approximation, respectively. Data adapted from Ref. \cite{malottki2021stability}.}%
\end{figure}
Here, we are investigating the eigenmodes of the ferromagnetic state, i. e. magnon excitations in a $50 \times 50$ hexagonal simulation box of the \PdfccFeIr{} system. In Fig. \ref{fig:fm_energy_dispersion}, the energy curves of the lowest three eigenmodes obtained by EMF are displayed up to the limit of $50$ meV/atom, which corresponds to an average temperature of $380$ K. All three energy curves exhibit a parabolic form and can be modelled well by the harmonic approximation as expected. The EMF method yields almost exactly the same energy curves which shows that it works for systems of this size and in the regime of the harmonic approximation. 

\begin{figure}
\includegraphics[width=84mm]{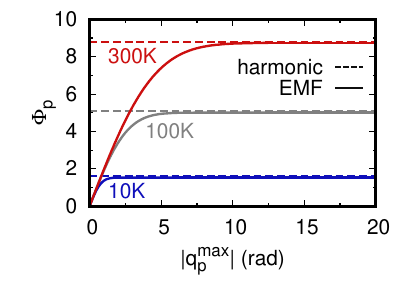}
\caption{\label{fig:magnon_mode_fm}%
\textbf{Value of $\Phi_p$ of the first magnon mode over the maximal EMF integration distance, $q_p^{max}$.} %
The numerical integration of the Boltzmann factor has been performed for three different temperatures and the corresponding integrals solved analytically via the harmonic approximation are shown as dashed lines. Figure adapted from Ref. \cite{malottki2021stability}.%
}
\end{figure}
To test the cutoff parameter of the numerical integration, $q_p^{\text{max}}$, the partition function, $\Phi_p$, of the first magnon mode has been calculated with different values of the maximal displacement (see Fig. \ref{fig:magnon_mode_fm}). Note that the mode has been followed in positive and negative direction of the eigenvector until $|q_p^{\text{max}}|$ is reached in both directions and the contributions are added to yield $\Phi_p$.

In Fig. \ref{fig:magnon_mode_fm}, the values of $\Phi_p$ evaluated with EMF are converging with increasing $|q_p^{\text{max}}|$ against the corresponding dashed lines of the harmonic approximations. It is evident, that partition functions of lower temperatures are converging faster as their respective Boltzmann factors decline more quickly with increasing excitation energy. Overall, Fig. \ref{fig:magnon_mode_fm} demonstrates, that the evaluation of partition functions with EMF converges fast enough and in a displacement interval that is realistically feasible. 

%-------------------------------------
\section*{EMF Applications for magnetic skyrmions}
\label{sec:applications}
%-------------------------------------
%
While we have developed the EMF method for direct entropy calculation in the framework of atomistic spin simulations with a focus on magnetic skyrmions (see Ref. \cite{malottki2021stability}), we believe that the concept of the method is applicable in other fields as well. However, we found several significant use-cases in skyrmionics of which three are presented in this section. 

In all three examples, EMF was used to visualise and characterise the studied eigenmodes, yielding a deeper understanding of the physics at play. Even more importantly, however, does the EMF method allow for the calculation of partition functions and thus entropies of anharmonic eigenmodes which traditionally cannot be modelled adequately with the zero or harmonic approximations. This has been leveraged in the examples A and B. Further, EMF grants access to intermediate regimes and thus enables systematic parameter studies in which the characters of the eigenmodes change continuously with interaction parameters (example A and C) or temperature (example B).

\subsection*{A: Helicity degree of freedom}
\begin{figure}[!h]
\includegraphics[width=0.5\textwidth]{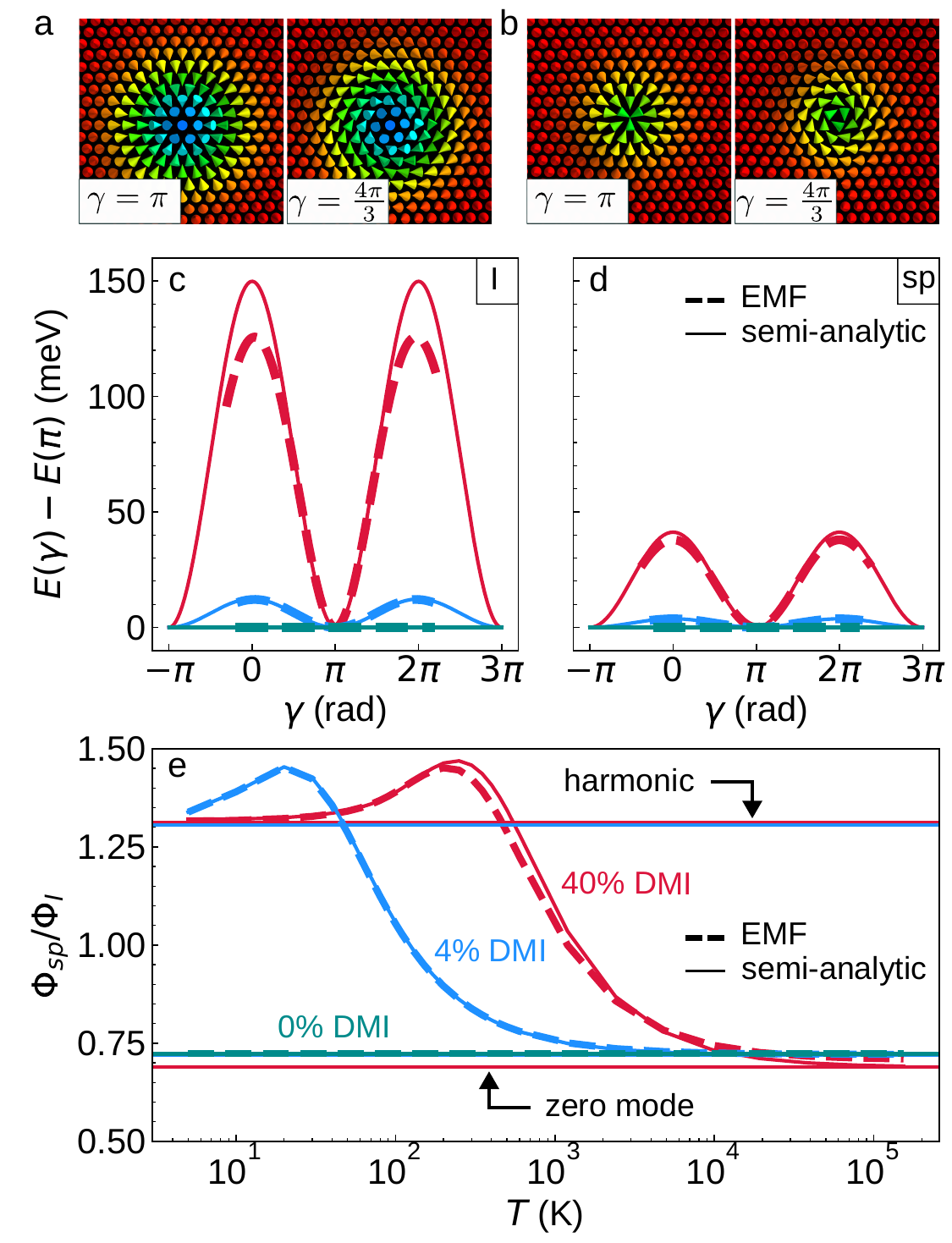}
\caption{\label{fig:helicity}
\textbf{Helicity degree of freedom for skyrmions}. Skyrmion (a) and radial saddle point (b) are shown for two helicities $\gamma=\pi,~\frac{4\pi}{3}$. The energy along the rotation of helicity for different DMI strengths is displayed in (c) for the skyrmion and in (d) for the saddle point. The dashed lines give the numeric results of the EMF algorithm while solid lines show semi-analytic results. The ratio of partition functions of skyrmion and saddle point states are presented in (e) for a range of temperatures and compared to zero and harmonic mode approximation.}
\end{figure}

\begin{figure*}[t]
\includegraphics[width=\textwidth]{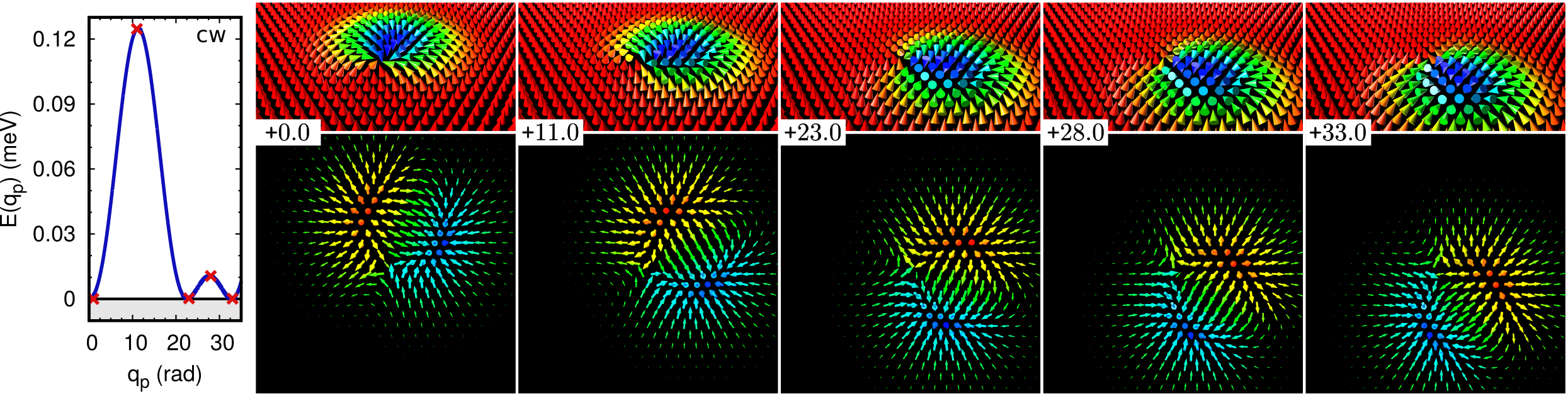}
\caption{\label{fig:chimera_rotation_cw_classification}
\textbf{Second chimera saddle point mode analysed by EMF.} The energy dispersion over the total geodesic distance, $q_p$, of a $2\pi/3$ rotation is shown on the left. For the marked positions the magnetic structures and the corresponding eigenvectors are shown on the right. Figure taken from Ref. \cite{malottki2021stability}.%
}
\end{figure*}

The helicity degree of freedom of skyrmions \cite{lin2016ginzburg} describes the collective rotation by an angle $\gamma$, also called helicity \cite{nagaosa2013topological}, of all spins in a 2d-system around an axis perpendicular to the film plane as illustrated in Fig.~\ref{fig:helicity} (a) for an arbitrary skyrmion state. It therefore translates between the Néel- ($\gamma=0,\pi$) and Bloch-configuration ($\gamma=\pm\pi/2$) of skyrmions. Because of this rather simple behaviour the trajectory of the corresponding mode can be parameterised by $\gamma\in[0,~2\pi]$ so that the orientation of every spin $\mathbf{m}_n$ can be described as
\begin{equation}\label{Eq:helicty_trajectory}
    \mathbf{m}_n(\gamma)=R_z^{\gamma-\gamma_0}\mathbf{m}_n(\gamma_0)
\end{equation}
where $\gamma_0$ is the helicity of the initial configuration and $R_z^{\gamma-\gamma_0}\in\mathbb{R}^{3\times3}$ describes a rotation of $\gamma-\gamma_0$ around the $z$-axis. This parameterisation gives rise to the zero mode volume of the helicity mode\cite{goerzen2023lifetime},
\begin{equation}
    V = 2\pi\sqrt{\sum_{n=1}^N 1-\left[m^z_n(\gamma_0)\right]^2}~.
\end{equation}
Within the extended Heisenberg model (Eq.~\ref{eq:heisenberg_model}) the uniform rotation around the $z$-axis leaves the exchange and Zeeman interaction as well as the magnetocrystalline anisotropy energy invariant. Only the DMI $E_{\text{DMI}}(\gamma) = E_{\text{DMI}}(\gamma + 2\pi)$ will undergo periodic modulations whose amplitude depends on the strength of the DMI.

As a parameter study, we relaxed skyrmions in the ultrathin Rh/Co/Ir(111) system at zero field \cite{meyer2019isolated, goerzen2023lifetime} and varied the strength of the DMI to $0\%, 4\%$ and $40\%$ of the original values given in Ref.~\cite{meyer2019isolated}. The calculations have been performed for skyrmions in simulation box with 50$\times$50 lattice sites, which have been minimized with respect to their energy by an VPO algorithm. The SP of radial skyrmion collapse have been calculated with CI-GNEB \cite{bessarab2015method} and illustrated in Fig.~\ref{fig:helicity}~b. We obtain the energy curves along the rotation for both, the skyrmions (Fig.~\ref{fig:helicity}~c) and the SPs (Fig.~\ref{fig:helicity}~d) by following the rotation eigenmode with EMF as well as a semi-analytical method that applies the parameterisation of Eq.~\ref{Eq:helicty_trajectory} in rigid-profile approximation. The periodic modulation due to DMI is evident for both structures and both, the EMF and the semi-analytical method yield similar results for all but the $40\%$ DMI curve. In this case, the energy increases obtained with EMF are lower than the semi-analytical one due to deformations of the radial skyrmion and SP profiles that the rigid-profile approximation cannot capture. 

The ratio of the partition functions, $\Phi^{SP}/\Phi^{I}$, which is an important part of the prefactor (Eq.~\ref{Eq:prefactor_general}), is displayed in Fig.~\ref{fig:helicity}~e over the temperature. The two methods, EMF and semi-analytic, yield similar curves besides the already discussed small deviation for $40\%$ DMI. For $0\%$ DMI, the ratio of partition functions is constant, reflecting the zero mode behaviour of the rotation eigenmode in absence of DMI. For finite DMI, the character of the eigenmode depends on the temperature. At $4\%$ DMI, it is harmonic only at $T<5$K. For larger temperatures, the ratio increases over the value of the harmonic approximation before it drops and converges to the value of the zero mode approximation between $10^3$ and $10^4$ K. For $40\%$ DMI, the curve is approximately shifted on the temperature axis by an order of magnitude. This highlights, that an adequate analytical approximation of the energy curve of eigenmodes depends strongly on the temperature while the EMF method is capable of describing the rotation mode over all displayed temperatures without prior knowledge of the nature or symmetry of the mode. Consequently, the EMF method has been used to calculate entropies and mean lifetimes of skyrmions and antiskyrmions in Rh/Co/Ir(111) as published in Ref. \cite{goerzen2023lifetime}.

\subsection*{B: Rotation of chimera mechanism saddle point}

The dominating mechanism of magnetic skyrmions annihilation depends on the material parameters and external stimuli \cite{meyer2019isolated, muckel2021experimental, malottki2021stability, desplat2019paths}. In case of a large DMI, a strong exchange frustration or effects that break the in-plane symmetry of the skyrmion, such as an in-plane magnetic field, skyrmions tend to decay via the so-called chimera collapse mechanism \cite{muckel2021experimental}. After its theoretical discovery \cite{desplat2019paths, meyer2019isolated}, the calculation of its entropic contribution to stability was prevented by the inability to handle the second eigenmode of the chimera transition state. 

\begin{figure}
\includegraphics[width=80mm]{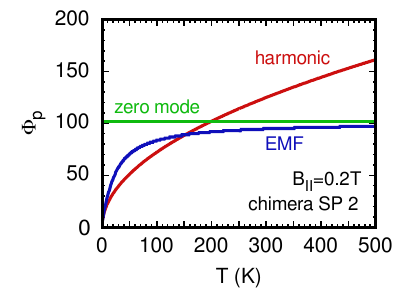}
\caption{\label{fig:chimera_mode_sp2_phi}
\textbf{Comparison of $\Phi_p$ of the second chimera saddle point eigenmode over temperature.} %
The values are obtained  numerically via EMF (blue), harmonic (red) or zero-mode approximation (green) for an external field of $B_\perp = 4.0T$ and $B_{||} = 0.2T$ in \PdfccFeIr{}. Figure adapted from Ref. \cite{malottki2021stability}.%
}
\end{figure}
\begin{figure*}[!t]
\includegraphics[width=0.7\textwidth]{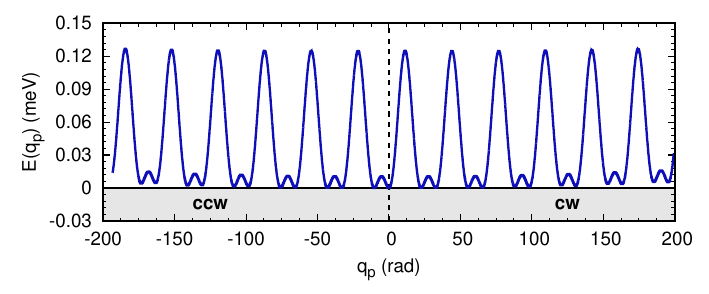}
\caption{\label{fig:dispersion_long_range}
\textbf{The energy over the distance $q_p$ along the second eigenmode of the chimera saddle point state.} The eigenmode has been followed with EMF in clock-wise (cw) and counter-clock-wise (ccw) rotational sense for several periods. Figure taken from Ref. \cite{malottki2021stability}.%
}
\end{figure*}

The application of the EMF method to the second eigenmode of the chimera SP (the first one is the unstable mode) reveals a rotation mode as illustrated on the right side of Fig.~\ref{fig:chimera_rotation_cw_classification}. On the left side, one period of the energy curve along the rotation is shown, which corresponds to a third full rotation of the Chimera SP structure. It is caused by the DMI, halving the six-fold symmetry of the hexagonal lattice structure into a 3-fold rotation symmetry. The energy differences on the order of 0.1 meV are extremely small, which makes them negligible at room temperature and this eigenmode suitable for a zero-mode approximation as long as its trajectory can be modelled accordingly. 

When applying an in-plane magnetic field to the spin structure, the 3-fold symmetry is broken, creating a preferred orientation of the moments and introduces an energy barrier for rotations with increasing magnetic field strength (see supplementary Fig. 1). For $B_{||} = 0.2$ T, this barrier is on the order of several meV, making the correct approximation of the energy curve dependent on the assumed temperature. In Fig.~\ref{fig:chimera_mode_sp2_phi}, the contribution to the partition function, $\phi_p$, has been determined for different temperatures using a zero-mode approximation, EMF and an harmonic approximation, respectively. The zero-mode value is constant as its occupation is independent of the temperature, while, on the contrary, the harmonic approximation assumes an ever increasing occupation. The EMF method produces a physically more sound overall behaviour, aligning for low temperature with a more harmonic behaviour and then gradually converging towards the value of the zero-mode approximation. This example demonstrates, that the accurate calculation of partition functions in the intermediate temperature regime is feasible while the traditional zero and harmonic approximation fail. 

In order to test the numerical stability of the EMF method, we perform long distance EMF calculations, following the rotational chimera SP mode for several full rotations. The resulting the energy curves for clockwise and counter-clockwise rotation are shown in Fig.~\ref{fig:dispersion_long_range}. When reaching the next local energy minimum, the obtained spin structures along the EMF exhibit the same energy as the initial ground state. However, after more than three full rotations ($|q_p|>|\pm 100|$ rad), small but increasing deviations of the energy are becoming visible, indicating increasing deviations in the spin structure due to numerical errors. This demonstrates that the numerical robustness of the EMF method, even in the simple Euler implementation, goes far beyond linear response approaches and is sufficient for the calculation of one period of the energy curve along the chimera SP rotation eigenmode. Consequently, the EMF method has been applied to calculate the lifetimes of skyrmions collapsing via the chimera mechanism of which further details can be found in Ref. \cite{muckel2021experimental, malottki2021stability}.

\subsection*{C: Exploration of reaction mechanisms}
\label{ssec:exploration_appl}
While the CI-GNEB method yields the minimum energy path and the SP(s) along that path, the EMF method enables the exploration of the PES perpendicular to the path along selected eigenmodes. Here, we want to demonstrate how such an application of the EMF can be helpful in understanding the physics of degenerate reaction mechanisms that change with varying system parameters. For this purpose, we consider a toy model of four spins arranged in a square, interacting via
\begin{equation}
    \begin{split}\label{eq:4spin_hamiltonian}
        E &= -J(\mathbf{m}_1\cdot\mathbf{m}_2+\mathbf{m}_1\cdot\mathbf{m}_3+\mathbf{m}_2\cdot\mathbf{m}_4+\mathbf{m}_3\cdot\mathbf{m}_4)\\
        &-\sum_{i=1}^4 K(\mathbf{m}_i\cdot\hat{\mathbf{z}})^2~.
    \end{split}
\end{equation}
Here, $J$ is the nearest neighbour exchange coupling parameter and $K=1.0$~meV is the uniaxial magnetocrystalline anisotropy. In order to flip from a configuration with all spins upwards to one with all spins downwards, it is intuitively clear that all spins will rotate simultaneously for strong exchange coupling of $J\gg 1$~meV. For intermediate coupling, the spins should flip pairwise and for very weak coupling they should flip individually.

\begin{figure*}
\includegraphics[width=\textwidth]{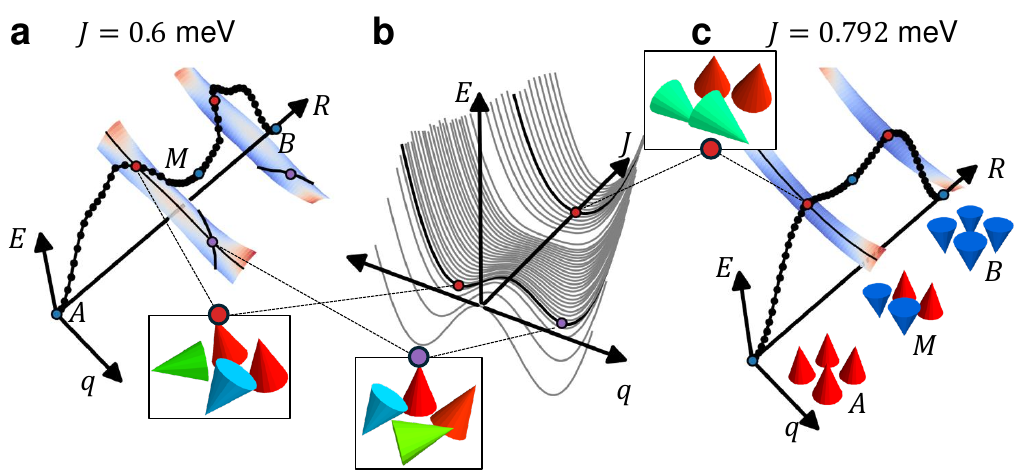}
\caption{\label{fig:application_4spin}
\textbf{Exploration of potential energy surface to reveal a Landau transition.} %
a,c: Minimum energy path between the two degenerate ground states $A$ and $B$ of the Hamiltonian in Eq.~\ref{eq:4spin_hamiltonian} for two different parameters of the exchange interaction. The energy landscape perpendicular to the reaction coordinate was determined along the third lowest eigenmode using the EMF method for the images in the vicinity of the saddle points (red). The respective degenerate saddle point configurations are shown in purple. b: Energy landscape over the travelled geodesic distance $q$ along the third lowest mode starting from the saddle point for different values of the exchange parameter $J$. 
}
\end{figure*}

The actual MEPs of the toy problem obtained by CI-GNEB are displayed in Fig.~\ref{fig:application_4spin}~a,c. For $J=0.6$~meV, there are two degenerate MEPs, of which one is shown. They correspond to the weak coupling regime in which each spin flips individually, although the coupling is already strong enough to affect the other spins, too. This leads to two MEPs of sequential pairwise rotation. By applying EMF to the third eigenmode of the SP state (the first and second being the unstable and a zero mode), one can move from one SP (red dot) to the other (violet dot). The resulting energy dispersions along that mode are  illustrated by the stripes of PES perpendicular to the MEP for the first and second SP. For the stronger coupling of $J=0.792$~meV, the situation changes to a single MEP of simultaneous pairwise rotation.

Now, applying this combination of GNEB and EMF systematically, the dependency of the energy dispersion of the third eigenmode for varying values of $J$ can be visualised (Fig.~\ref{fig:application_4spin}~b). It can be seen how the system undergoes a Landau transition from two degenerate SPs to one. Note that Fig.~\ref{fig:application_4spin}~b only shows the first SP of each MEP although a similar plot could also be created for each second SP. This application of the EMF method allows to analyse the underlying physics of a transition in reaction mechanisms qualitatively as well as quantitatively. This procedure has been applied to study the behaviour of \PdfccFeIr{} bilayers coupled by varying strength of the interlayer exchange interaction as published in Ref.~\cite{schrautzer2022}. Furthermore, we expect that exploration using the EMF perpendicular to the reaction coordinate is useful in systems with post transition state bifuractions~\cite{metiu1974,valtazanos1986,quapp1998bifurcation,garcia2021}.

\section*{Conclusion}
In this work we showed how the EMF method can be used to calculate partition functions and thus entropies in the framework of transition state theory numerically, going beyond the zero mode and harmonic approximations. The algorithm requires the iterative calculation of eigenpairs and a continuous potential energy surface allowing for successful mode tracking. 

In the framework of atomistic spin simulations, the method has been illustrated on the example of a single spin affected by uniaxial magnetocrystalline anisotropy, highlighting that it can be applied in both, the regime of the zero-mode approximation as well as the regime of the harmonic approximation. On the example of magnon excitations in a ferromagnetic state, the convergence of the numerical integration of the partition functions with feasible cutoff parameters, $q_p^{\text{max}}$, and how these parameters depend on temperature have been demonstrated.

The use-case examples focusing on skyrmion collapse showcased that partition functions and therefore entropies and prefactors in relevant and complex systems with initially unknown and anharmonic eigenmodes can be calculated with the EMF method. The 4 spin toy model with changing transition mechanisms under varying exchange coupling, $J$, showed how EMF can be of use for understanding and analysing the underlying physics of an eigenmode as it has been applied for realistic multilayer systems. Moreover, the flexible nature of EMF enables systematic parameter studies beyond the conventional boundaries of specific approximations as demonstrated with the DMI variation in the Rh/Co/Ir(111) system in which the rotational eigenmode continuously changed its character from a zero to an harmonic eigenmode. Besides its direct application to determine entropy contributions, EMF is also a powerful tool to examine the nature of eigenmodes in the first place and to evaluate the quality of conventional methods which have been often used without any information about their applicability.

The EMF method for direct entropy calculation does not require detailed prior knowledge of the system or eigenmodes and is flexible in the implementation of the eigenpair calculation, making it a very general approach to obtain entropies beyond harmonic approximation. In quasi-classical systems like atomistic spin simulations, this allows a straightforward treatment of softest eigenmodes. In large systems, it is good practice to combine the accuracy of EMF with the efficiency of the harmonic approximation, allowing for entropy and prefactor calculations of realistic systems. While it remains to be seen if the EMF method can be transferred directly to quantum mechanical methods such as DFT, we believe that it can be useful in a wide range of fields in physics and chemistry.

\section*{Methods}
\subsection*{Simulation details}
All atomistic spin simulations have been performed with Spinaker and its predecessor, the Kiel simulation code. Both codes were developed in the collaboration of the University of Kiel and the University of Iceland. The EMF method is implemented with a simple Euler rotation step, realizing a geodesic displacement along the selected eigenvector. The EMF calculations typically consist of $10000$ to $20000$ of such steps with a constant width of $0.5-1.0~*~10^{-4}$ rad. For the exploration of the energy landscape perpendicular to the reaction coordinate (Fig.~\ref{fig:application_4spin}~b) steps with a constant width of $0.025$~rad are used. Magnetic textures were energetically relaxed via spin-dynamics, vpo or the lbfgs algorithm with applied periodic boundary conditions. MEPs and SPs have been obtained via CI-GNEB.

\subsection*{Calculations of the partition functions of a single spin with various methods}
Here we provide the details on how the values in Tab. \ref{Tab:phi_single_spin} have been obtained. In a single spin system with uniaxial anisotropy along $\mf{\hat{z}}$, the energy is given by
\begin{align}
E\ &=\ -K\ m_z^2\\
&=\ -K\ \cos^2 (\theta).
\end{align}
The strength of the anisotropy is $K = 1~$meV and the resulting energy of the ground state is $E_0 = -1~$meV.

In the case of the spin being exactly aligned with the easy axis, there is a two-fold degenerate eigenmode describing the direct rotation from pole to pole. Its partition function is given by
\begin{align}
\Phi_{\text{min}, 1} &= \int e^{-\beta [E(\theta) - E_0]}~ d\theta\\
&= \int e^{-\beta K [\cos^2(\theta)-1]}~ d\theta
\label{eq:min_1_integral}
\end{align} In harmonic approximation and with the eigenvalue of the Hessian, $\lambda_1 = 2~$meV, the integral simplifies to
\begin{align}
    \Phi_{\text{min}, 1}^{\text{harm}} &= 2.0 * \sqrt{\frac{2 \pi}{\beta \lambda_1}}\\
    &\approx 3.291~.
\end{align}
The combinational factor of two stems from the two local minima at $\pm \hat{\mf{z}}$ which are approximated by one parabola each and both contribute to the occupied states when considering a full rotation.

In zero mode approximation, the integral reads
\begin{align}
    \Phi_{\text{min}, 1}^{\text{zero}} &= \int^{2\pi}_{0} e^{-\beta [E(\theta = 0) - E_0]} ~ d\theta\\
    &= \int^{2\pi}_{0} 1 ~ d\theta\\
    &= 2\pi
\end{align}
For a solution without approximating of the energy function, we use equation \ref{eq:min_1_integral} and solve it by direct numerical integration over $\theta$. This is possible because the trajectory of the eigenmode is trivial and known: A straight rotation from $+\hat{\mathbf{z}}$ to $-\hat{\mathbf{z}}$ and back on the other side of the sphere. For the application of the EMF method, the initial spin orientation is very slightly rotated away from the easy axis to avoid the initial degeneracy. However, this is done for numerical reasons and the affect on the result is highly negligible.

In the case of the spin being in SP configuration, i. e. it lies in the equator plane, the diagonalization of the Hessian yields two distinct eigenmodes. The first is similar in character to the two-fold degenerated mode of the energy minimum state we just discussed. The second corresponds to a rotation around the equator. Both eigenmodes cannot be integrated in harmonic approximation: The eigenvalue of the first eigenmode is negative (-2 meV) and the second eigenvalue is zero. To calculate them in zero mode approximation, we assume that the system has the constant energy of the saddle point state, $E_0^{\textnormal{SP}} = E(\theta = \pi/2) = 0.0~$meV, which is a bad approximation for the first eigenmode but an excellent one for the second.

For the first SP eigenmode in zero mode approximation, we get
\begin{align}
    \Phi_{\text{SP}, 1}^{\text{zero}} &= \int^{2\pi}_{0} e^{-\beta [E(\theta=\pi/2)-E_0^{\textnormal{SP}}}] ~ d\theta\\
    &= 2\pi~ e^{\beta E_0^{\textnormal{SP}}}\\
    &\approx 1.969~.
\end{align}
The second SP eigenmode in zero mode approximation yields the same value as we just integrate the same constant over a full period over $\phi$ instead of $\theta$:
\begin{align}
    \Phi_{\text{SP}, 2}^{\text{zero}} &= \int^{2\pi}_{0} e^{-\beta [E(\phi=\pi/2) -E_0^{\textnormal{SP}}]} ~ d\phi\\
    &= 2\pi~ e^{\beta E_0^{\textnormal{SP}}}\\
    &\approx 1.969~.
\end{align}
Both eigenmodes can be solved numerically by applying the direct numerical integrations of $\theta$ and $\phi$, respectively, whose trajectories are trivial. The EMF method can be applied without any special considerations.

\section*{Data Availability}
The data that support the ﬁndings of this study are available from the corresponding author upon reasonable request.

\bibliography{references}

\section*{Acknowledgements}
S.v.M. gratefully acknowledges funding of the Deutsche Forschungsgemeinschaft (DFG) via a Walter Benjamin-Stipendium with project number 523127890 as well as funding from the Fonds National de la Recherche Scientifique (FNRS) under project number 1.B.504.24. H.S. gratefully acknowledges financial support from the Icelandic Research Fund (Grant No. 239435). P.F.B. gratefully acknowledges financial support from the Icelandic Research Fund (Grants No. 217750 and 2410333), the University of Iceland Research Fund (Grant No. 15673), the Swedish Research Council (Grant No. 2020-05110), and the Crafoord Foundation (Grant No. 20231063). M.A.G. and S.H. 
gratefully acknowledge financial support from the DFG
through SPP2137 “Skyrmionics” (Project No. 462602351).
We thank Dr. Johannes Steinmetzer and Prof. Henrik Grönbeck for helpful discussions. 

\section*{Author contributions}
S.v.M., M.A.G. and H.S. performed the calculations. S.v.M. conceived the project and developed the EMF method and its original implementation in the Kiel simulation code. M.A.G. optimized the implementation. H.S. implemented the EMF method in the Spinaker code and further optimized it. All authors contributed to the analysis and discussion of the simulation results. S.v.M., M.A.G. and H.S. wrote the first version of the manuscript, and all authors contributed to the final version. 

\section*{Competing interests}\noindent
The authors declare no competing financial or non-financial interests.

\section*{Additional information}
\subsection*{Supplementary material}\noindent
This publication is accompanied by supplementary material.
\subsection*{Correspondence}\noindent and requests for further details should be addressed to Stephan von Malottki

\end{document}